\input epsf

\documentclass[prb,aps,psfig,twocolumn]{revtex4}
\usepackage{graphicx}
\usepackage{afterpage}

\begin{document}
\title{Escaping free-energy minima}
\author{Alessandro Laio and Michele Parrinello}
\affiliation{CSCS--Centro Svizzero di Calcolo Scientifico, Via Cantonale, CH-6928 Manno, Switzerland \\
Department of Chemistry, ETH, H\"{o}nggerberg HCI, CH-8093 Zurich, Switzerland}
\date{\today}

\begin{abstract}
We introduce a novel and powerful method for exploring the properties of the multidimensional
free energy surfaces of complex many-body systems by means of a coarse-grained 
non-Markovian dynamics in the space defined by a few collective coordinates.
A characteristic feature of this dynamics is the presence of a history-dependent 
potential term that, in time, fills the minima in the free energy surface, 
allowing the efficient exploration and  accurate determination of the free energy 
surface as a function of the collective coordinates.  We demonstrate the usefulness of this 
approach in the case of the dissociation of a NaCl molecule in water and in the study 
of the conformational changes of a dialanine in solution.
\end{abstract}

\maketitle

Molecular dynamics and the Monte Carlo simulation method have had a very deep 
influence on the most diverse fields, from materials science to biology, and from 
astrophysics to pharmacology.  Yet in spite of their success, these simulation methods 
suffer from limitations that reduce the scope of their applications.  A severe constraint 
is the limited time scale that present-day computer technology and sampling 
algorithms explore.  In particular, there are many circumstances where the free energy 
surface (FES) has several local minima separated by large barriers.  Examples 
of these situations include conformational changes in solution, protein folding, first 
order phase transitions, and chemical reactions.  In such circumstances a simulation 
started in one minimum will be able to move spontaneously to the next minimum only 
under very favourable circumstances.  A host of methods have been suggested in 
order to lift this restriction and to explore the FES
\cite{Constraints1,Huber,Grubmuller,Constraints2,Voter1,Voter2,Steiner,Gong,VandeVondele1,free_energy,wang,VandeVondele2,rahman}, 
or to characterize the transition state \cite{chandler,passero}.
Here we find a new solution to this problem by combining the ideas of
coarse-grained dynamics\cite{kevrikidis,E} on the FES\cite{free_energy,VandeVondele2} with those of adaptive bias
potential methods\cite{Huber,wang} obtaining a procedure that allows the system to
escape from local minima in the FES and, at the same time,
permits a quantitative determination of the FES as a by-product of the
integrated process.

\section{Methodology}
We shall assume here that there exist a finite number of relevant collective 
coordinates   $s_i, \,\, i=1,n $   where $n$ is a small number, 
and we consider the dependence of 
the free energy  ${\cal F}(s)$ on these parameters.  
Practical examples of appropriate choices of these variables will be given below.  
The  exploration of the FES  is guided by the forces 
$F^t_i=- {\partial {\cal F}} / {\partial s^t_i}$.
In order to estimate these forces efficiently, we introduce an ensemble 
of $P$  replicas of the system, 
each obeying the constraint that the collective coordinates have a pre-assigned value 
$s_i=s_i^t$, and each evolved independently at the same temperature $T$.  
Since the $P$ replicas are statistically independent, the estimate of thermodynamic
observables (e.g. the forces on the constraints) is improved with respect to an evaluation
on a single replica, and it can be parallelized in a straightforward manner.
The constraints are imposed on each replica via the standard methods 
of constrained molecular dynamics\cite{allen} by adding to the Lagrangean a term 
$\sum_{i=1,n} \lambda _i (s_i-s_i^t)$, where $\lambda _i$ are Lagrange multipliers.
Averaging over the time and over the   replicas 
we can evaluate the derivative of the free energy relative to the  $s_i^t$'s as 
$F^t_i=\left <  \lambda _i \right >_t $\cite{Constraints1,Constraints2}.  
For the sake 
of simplicity we neglect here the small kinematic correction term discussed in Ref. \cite{Constraints2}, 
although this can easily be added.  Taking a cue from the work of Gear and 
Kevrikidis\cite{kevrikidis}, we use these forces, determined by a  microscopic dynamics, 
to define a coarse-grained 
dynamics in the space of the  $s_i$'s.  In our case the definition of this dynamics is rather 
arbitrary and designed only to explore the FES efficiently.  The dynamics is 
defined from the discretized evolution equation:
\begin{equation}
\sigma^{t+1}_i=\sigma^{t}_i+\delta \sigma \frac {\phi_i^t} {\left| \phi^t \right|}
\label{dyn}
\end{equation}
In eqn. (\ref{dyn}) we have introduced the scaled variables $\sigma^t_i=s^t_i / \Delta s_i $ and 
the scaled forces $\phi_i^t = F_i^t  \Delta s_i $,  
where $\Delta s_i$ is the 
estimated size of the FES well in the   direction $s_i$, $\left| \phi^t \right| $ 
is the modulus of the 
 n-th dimensional vector  $\phi_i^t$ and $\delta \sigma$ is a  dimensionless stepping parameter.  
It should be stressed that 
eqn.  (\ref{dyn}) has the form of a steepest descent equation in the direction given by the forces
$\phi_i$, and  does not imply any real dynamical evolution. The index  $t$ is only used to 
label the states.  After the collective coordinates are updated using eqn.  (\ref{dyn}), a new 
ensemble of   replicas of the system with values  $\sigma^{t+1}_i$ is prepared, and new forces 
$F^{t+1}_i$ are calculated for the next iteration.  
At the same time the driving forces 
are evaluated from the microscopic Hamiltonian in short standard microscopic 
molecular dynamics runs. 
Since there is no dynamical continuity 
between the $P$ replicas at different iterations, one can use large values of $\delta \sigma$
and move very efficiently in the space of the collective coordinates. 

Clearly eqn. (\ref{dyn}) alone cannot guarantee an efficient exploration of the FES, nor is it 
useful to determine the FES.  
This task can be achieved if we replace the forces in 
eqn. (\ref{dyn}) with a history-dependent term\cite{Huber,wang}:

\begin{equation}
\phi_i \rightarrow \phi_i - \frac \partial {\partial \sigma_i} W
\sum_{t' \le t} \prod _i e^{ -\frac {|\sigma_i-\sigma^{t'}_i|^2}{2 \delta \sigma^2} }
\label{gauss}
\end{equation}
where the height and the width of the Gaussian W 
and $\delta \sigma$ are chosen in order to provide a reasonable compromise between accuracy and
efficiency in exploring the FES, as we will show below.
The component of the forces coming from the Gaussian will discourage the system from 
revisiting the same spot 
and encourage an efficient exploration of the FES.  As the system diffuses though the 
FES, the Gaussian potentials accumulate and fill the FES well, allowing the system to 
migrate from well to well.  After a while the sum of the Gaussian terms 
  will almost exactly compensate the underlying FES well. 

\begin{figure}[htb]
\setlength{\abovecaptionskip}{0cm}
\centering
\centerline{\includegraphics[width=7.7cm,angle= 270,clip=]{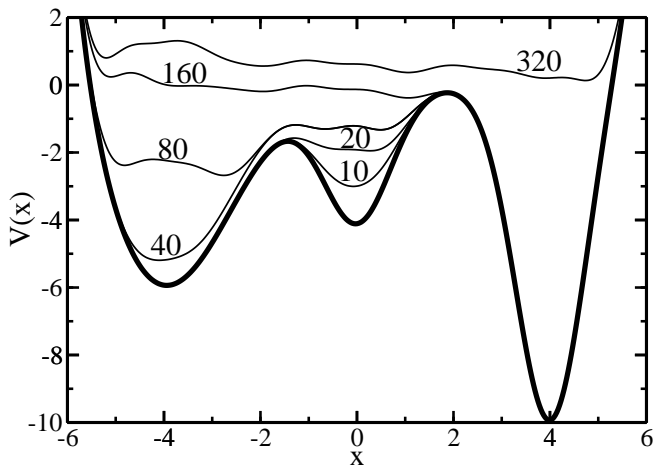}}
\renewcommand{\baselinestretch}{1.0}
\caption{ 
Time evolution of the sum of a one-dimensional model potential V($\sigma$) and 
the accumulating Gaussian terms of eqn. (\ref{gauss}). The dynamic evolution (thin lines) 
is labelled by the number of dynamical iterations (\ref{dyn}).
The starting potential (thick line) has three minima and the dynamics is initiated in the 
second minimum.
}
\end{figure}

 A typical example of this behaviour is shown in Figure 1, in which a dynamics in eqn. (\ref{dyn})
is used to explore a one-dimensional potential energy surface (PES) $V(\sigma)$ with three  wells. 
The dynamics starts from a local minimum that is filled by the Gaussians  in $\sim$ 20 steps. Then
the dynamics escapes from the well from the lowest energy saddle point, filling the second well in $\sim$ 80
steps. The second highest saddle point is reached in $\sim$ 160 steps, and the full PES is filled 
in a total of $\sim$ 320 steps.
Hence, in the case of this example, since the form of the 
potential is known, it can be verified that, for large t and if the width of the Gaussians 
is sufficiently small with respect to the length of a typical variation of V,
\begin{equation}
 -\sum_{t' \le t} W 
e^{ -\frac {|\sigma-\sigma{t'}|^2}{2 \delta \sigma^2} }  \rightarrow V (s) 
\label{FES}
\end{equation}
modulus an additive  constant.
The same property can be verified for any multidimensional potential if the variables $\sigma_i$ are allowed
to vary in a finite region. 
Hence, we assume that eqn. (\ref{FES}) holds also if the method is used to explore the FES, and that the 
free energy can be estimated from eqn. (\ref{FES}) for large t.
This procedure resembles the algorithm recently proposed by Wang and Landau\cite{wang}, in which a time-dependent bias
is introduced  in order to modify the density of states to produce locally flat hystograms.

We observe an interesting isomorphism between our dynamics and the
ordered deposition of the beads of a polymer chain on the FES. In fact we
can regard $\sigma _{i}^{t}$ as the position of a monomer in the
n-dimensional parameter space. Each monomer is held at a distance $\delta
\sigma $ from the previous one and monomers repell each other with the
Gaussian term of eqn. (\ref{gauss}). Neglecting terms of order $\delta \sigma ^{2}$
each monomer is placed by eqn. (\ref{dyn})  in the position of free enegy minimum compatible with
the requirement of minimum overlap with previous beads. Hence, the polymer chain generated 
in this manner fills the wells in the FES.  This isomorphism
helps to clarify how our approach works and why it can be made more precise
by reducing $\delta \sigma $.

The efficiency of the method in filling a well in the PES (or in the FES) can be estimated by the number of 
Gaussians that are needed to fill the well. 
This number is proportional to 
$ \left(  1 / \delta \sigma  \right)^{n} $, where n is the dimensionality of the problem. 
Hence, the efficiency of the method scales exponentially with the number of dimensions involved. 
If n is large, the only way to obtain reasonable efficiencies is to use Gaussians with a size
comparable to that of the well. On the other hand, a sum of Gaussians can only  
reproduce features of the FES on a scale larger than $\sim \delta \sigma$.
A judicious choice of $\Delta s_i $, W and $\delta \sigma$  will ensure the right 
compromise between accuracy and sampling efficiency,
and the optimal height and width of the Gaussians are determined by the typical variations of the FES.  

If prior information on the nature of the free-energy well is not
available, the scaling parameters $\Delta s_{i}$ are chosen by performing
short coarse-grained dynamic runs without bias potential (see  eqn. (\ref{dyn})). In
such a case the system moves around the minimum. The scaling parameters are
chosen so that the elongations  have approximately the same value in all
directions. This amounts to an empirical form of preconditioning which makes
the FES minimum nearly spherical in n dimensions and easy to fill with
n-dimensional Gaussians.

The metastep length $\delta \sigma $ determines the efficiency of the method
in a manner that is exponential in $n$ and it should be made as large as
possible. However, the dynamics in eqn. (\ref{dyn}) is able to reconstruct details of the FES  only on the
length scale of  $\delta \sigma $. Moreover, an over-large value of $\delta \sigma $ would cause the
system to jump irregularly from one basin to the other. 
The value of $\delta \sigma $ is chosen requiring 
the metatrajectory to remain localized in the FES minimum if the bias potential is not
applied.
With this choice of $\delta \sigma $, a single step of the dynamics in eqn. (\ref{dyn}) cannot 
lead a system from a FES minimum to another,
and the initial state at each new iteration can be generated from the last
MD step of the previous iteration without creating major overlaps.
Since the
shake algorithm is used to impose a new set of $\sigma _{i}^{t+1}$ values,
large forces on the constraints are generated at the beginning of each microscopic
dynamics, and the initial part of the trajectory has to be discarded for the calculation of $F_{i}^{t+1}$. 

The height of the Gaussians $W$ can be estimated as follows. We notice that we want to
reach a situation in which the modified FES is flat. In such a case the forces
coming from the FES and that coming from the Gaussian approximately balance
each other. If we require the maximum value of the Gaussian forces to be
smaller than the typical FES force, we arrive at the relation
$W / \delta \sigma \, e^{-\frac{1}{2}}=\alpha \left\langle
F_{i}^{2}\right\rangle ^{\frac{1}{2}}$ with $\alpha <1$. In all cases so far
studied a value of $\alpha $ close to $.5$ \ allows a fast escape from the local
minima in the FES without
significant loss of the underlying structure. 

In the future adaptive ways of determining all these parameters should be
considered in order to adapt the procedure in an optimal manner to the local shape of
the FES.

\begin{figure}[htb]
\setlength{\abovecaptionskip}{0cm}
\centering
\centerline{\includegraphics[width=7.7cm,clip=]{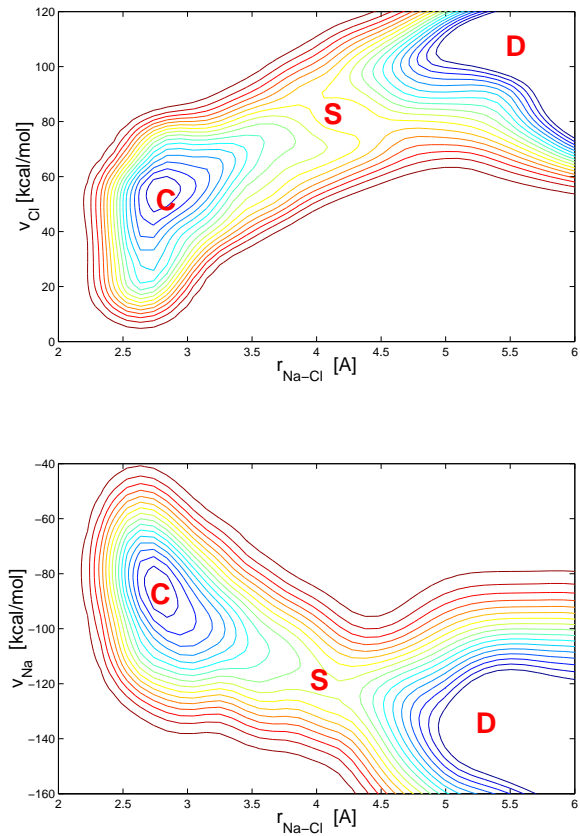}}
\renewcommand{\baselinestretch}{1.0}
\vspace{0.5truecm}
\caption{ 
Free energy as a function of the collective coordinates $r_{Na-Cl}$, $V_{Na}$ and $V_{Cl}$  
for a Na-Cl complex in water. The two-dimensional free energy F$_2$ 
shown here  are obtained as
$F_2=-1/\beta  \log \left[ \int dz \exp(-\beta F_3) \right] $, where $\beta$ is the 
inverse temperatures,  z is the coordinate not shown here, and the three-dimensional
free energy F$_3$ is estimated with eqn. (\ref{gauss}). 
The contours  are plotted every 0.5 kcal/mol. The zero of the 
free energy corresponds to the metastable minimum of F in the contact ion pair (C in
figure). 
}
\end{figure}

\section{Results}
\subsection{Dissociation of NaCl in water}
We first discuss an application of the method to the dissociation of 
a NaCl molecule in water.  The most stable state is the 
dissociated one, while the undissociated contact ion pair corresponds to a  metastable local minimum.
Here we shall study how the 
system escapes from this metastable state.  We model the system using the AMBER95 
force field\cite{amber}, and solvate the NaCl complex in 215 TIP3P water molecules with periodic 
boundary conditions. The
electrostatic interaction between classical atoms is taken into account by
the P3M method\cite{philippe}. As collective coordinates, 
we use the distance $r_{Na-Cl}$  between Na 
and Cl, and, in order to take into account the dynamics of the solvation shell during the dissociation,  
we also use the electric fields $V_{Na}$ and $V_{Cl}$ on the Na and on the Cl due to the water molecules 
within $\sim 6.5$ \AA $\,,$ of the ions. $V_{Na}$ and $V_{Cl}$ depend significantly on the hydration pattern
around the Na and on the Cl ion. If, for example, a hydrogen bond to one of the two ions is formed or broken, 
the field on the ion changes by several  kcal/mol.
A dynamics of the form (\ref{dyn}) was performed on the system, with $\delta \sigma=0.25$ and $W=0.3$ kcal/mol.
The scaling parameters 
$\Delta s_i$  were 0.53 \AA, 32 and 32 kcal/mol  for $r_{Na-Cl}$, $V_{Na}$ and $V_{Cl}$
respectively.
The forces in eqn. (\ref{dyn}) were evaluated by short MD runs with a time step of 0.7 fs 
performed on 6 replicas of the system.
The   replicas were then equilibrated for 200 MD steps at 300 K, and the forces  evaluated 
by averaging over the following 500 MD steps.  Starting from a distance of 2.6 \AA, corresponding to a
contact pair, the ions dissociate after 120 coarse-grained iterations. The dynamics was then 
continued for another 350 steps imposing a maximum separation of the ion pair of 5 \AA, in
order to study the structure of the free energy surface around the transition state (see Figure 2). 
Seven recrossings for the coarse-grained dynamics were observed.
The transition state is located $r_{Na-Cl} \sim 4.02$ \AA, $V_{Na} \sim -120$ kcal/mol and 
$V_{Cl} \sim 81.5$ kcal/mol (see Figure 2). The overall topology of the FES confirms the importance of
the solvent degrees of freedom for describing the reaction, since the transition region is in
transverse orientation relative to the axis of the coordinates\cite{geissler}, and hence neither the 
distance nor the fields alone can provide an exhaustive description of the dissociation.
The transition barrier (S in Figure 2), estimated with thermodynamic integration\cite{Constraints2}
along $r_{Na-Cl}$, is of 3.3 kcal/mol. The one-dimensional free energy as a
function of  $r_{Na-Cl}$ can also be obtained by integrating the three-dimensional FES ($F_3$) with
respect to $V_{Na}$ and $V_{Cl}$ as
$F_1=-1/\beta  \log \left[ \int dV_{Na} \,  dV_{Cl} \, \exp(-\beta F_3) \right] $.
This leads to a barrier of 3.4 kcal/mol.
Hence, the method is able to reproduce the one-dimensional free energy profile estimated with
thermodynamic integration.

 We have verified that these collective coordinates
identify the transition state and the reaction coordinate also in the dynamical sense\cite{chandler}.  To this effect we 
prepared a set of replicas of the system with the critical values of the collective coordinates. 
We then followed the procedure described in Ref.\cite{geissler},  assigning to the particles 
values of the velocity chosen from a Maxwellian 
distribution at 300 K, and following the trajectory backwards and forwards in time.
By this procedure, we found that, within a time scale of 
200 fs, the overwhelming majority of the replicas fell into one of the two 
basins of attraction without any recrossing. 
Moreover, the time-reversed dynamics
leads systematically to the opposite basin of attraction\cite{chandler}.
Finally, if we prepare the 
systems with the SHAKE algorithm\cite{allen} for a given value of the collective coordinates,  
and if the particle velocities are not reassigned  before 
letting the system evolve freely, the time needed to go in either of the attraction basins 
is of the order of a few ps, and several recrossings are observed.  
This is due to the fact that the SHAKE algorithm imposes 
zero velocities in the direction of the constraint and it is only after thermalization, 
which takes place in a time scale of picoseconds, that the collective coordinates acquire 
sufficient velocity to evolve towards the FES minima.  This is a strong confirmation 
that we have identified the correct transition state.

In order to characterize the solvent structure during the dissociation, we have computed the Na-Water
pair correlation function at the transition state S. The coordination number, obtained by integrating the 
pair correlation function up to a distance of 3 \AA, is 5.28. This value is consistent with the picture
depicted in Ref. \cite{geissler}, in which the sodium at the
transition state is shown to be typically only five-fold coordinated.

\subsection{Isomerization of alanine dipeptide in water}
As a second application, we used our method to explore the free energy surface of an
alanine dipeptide as a function of the backbone dihedral angles $\Phi$ and $\Psi$
\cite{bolhuis}. 
The system is described by the all-atom AMBER95 force field \cite{amber} and solvated in 
287 TIP3P water molecules. 
The parameters of the coarse-grained  dynamics of eqn. (\ref{dyn}) are $\delta \sigma=0.25$ and $W=0.25 $ kcal/mol, 
while the scaling parameters $\Delta s_i$ are $60^{o}$ for both $\Phi$ and $\Psi$.
The other settings are the same as for the simulation of the Na-Cl complex.
The FES of alanine dipeptide in water has been studied in detail by other authors\cite{bolhuis,apostolakis}, 
who identified several minima. 
Our dynamics  (\ref{dyn}) is able to capture the overall features of the FES extremely  quickly. 
\begin{figure}[h]
\setlength{\abovecaptionskip}{0cm}
\centering
\centerline{\includegraphics[width=7.2cm]{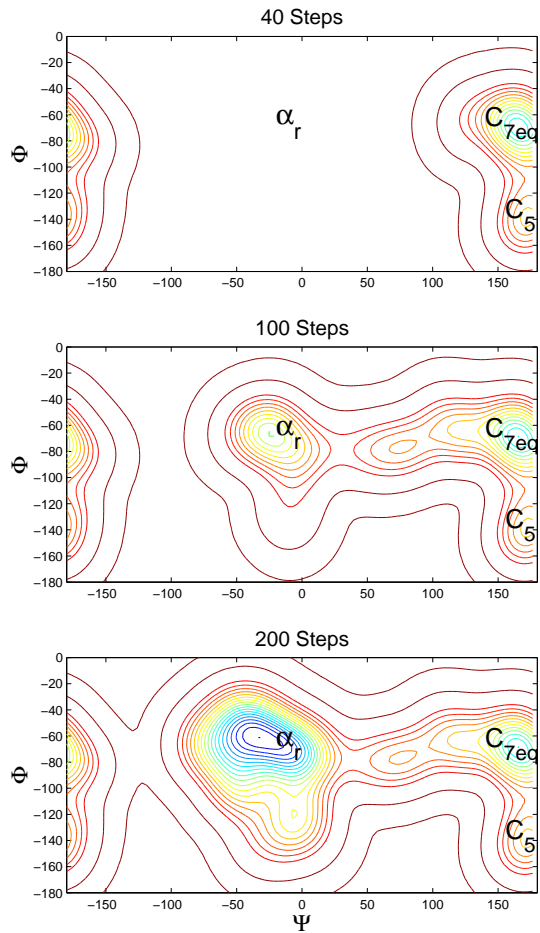}}
\renewcommand{\baselinestretch}{1.0}
\caption{ 
Free energy as a function of the backbone dihedral angles $\Phi$ and $\Psi$ for an
alanine dipeptide in water after 40, 100 and 200 steps of coarse-grained dynamics (\ref{dyn}).
The zero of the free energy corresponds to the the state $\alpha_r$, located at $\Phi=-63$ 
and $\Psi=-30$. The contours  are plotted
every 0.2 kcal/mol. The state $C_{7eq}$ is located at $\Phi=-70$
and $\Psi=165$ and 1.3  kcal/mol. Finally, the transition state between  
$\alpha_r$ and $C_{7eq}$  is located at $\Phi=-80$
and $\Psi=35$ and 3.3  kcal/mol.
If the dynamics is continued, less favourable areas of the phase space are explored (e.g. the region
with $\Phi>0$), but the 
free energy differences reported here are unchanged within the accuracy
of the calculation.
}
\end{figure}
Starting from a configuration close to the state $C_{7eq}$\cite{bolhuis,apostolakis}, the dynamics fills this attraction basin in approximately
45 steps. The minimum is localized at $\Phi \sim -80^o$ and $\Psi \sim 150^o$. The estimated free energy at
the saddle point is 2 kcal/mol higher than the free energy in $C_{7eq}$. This compares to a value of 2.1  
kcal/mol obtained in Ref. \cite{bolhuis} by thermodynamic integration. Our estimate 
is obtained with 45 steps of coarse-grained dynamics on 6 replicas, corresponding to a total of 
only $\sim 120 $ ps of mycroscopic dynamics. 
After another 150 steps, the dynamics fills the attraction basin of the state $\alpha_r$, which corresponds 
to the equilibrium configuration of the system, and recrosses to  $C_{7eq}$. Finally, 
the dynamics visits the $\Phi > 0$ region of the configuration space after $\sim 300 $ steps.
In Figure 3 we plot the free energy as a function of $\Phi$ and $\Psi$ after 40, 100 and 200 steps of 
coarse-grained dynamics. Also small details of the free energy landscape, like
the location of the saddle points and the presence of secondary minima (e.g. the state $C_5$
\cite{apostolakis}), can 
be detected graphically after a few steps of coarse-grained dynamics.
The one-dimensional free energy profile as a function of $\Psi$ obtained by analytic integration
of the FES reproduces the profile obtained in Ref. \cite{bolhuis} within 0.2 kcal/mol for all the values of 
 $\Psi$.

Concerning the choice of the collective coordinates for this system, it should be remarked that 
the dihedral angles $\Phi$ and $\Psi$ do {\it not} provide a complete 
description of the dialanine isomerization 
reaction\cite{bolhuis}, and the real reaction coordinate should
take into account also the solvent degrees of freedom. Still, our results show that
the method is able to reproduce the overall feature of the FES even if relevant degrees of freedom
are not explicitly included in the collective coordinate space. 
This shows that the neglected degrees of freedom, although relevant for determining the
reaction coordinate, are associated with relatively small barriers and are  sampled efficiently 
during the microscopic dynamics for each value of 
$\Phi$ and $\Psi$, in spite of the relatively short runs. This is due to the use of the replicas
and to the ability of the dynamics to retrace the  same region of parameter space during the dynamics.

\section{Discussion}
The method is based on the combination of  the ideas of coarse-grained dynamics\cite{kevrikidis}
in the space defined by a few collective coordinates, and on the introduction of a  history-dependent 
bias\cite{Huber,wang}.
Constructing a dynamics on a  free energy surface that depends on a few collective coordinates, 
we simplify enormously the complexity of the problem, which depends 
exponentially on the number of degrees of freedom. 
The FES is much smoother than 
the underlying potential energy surface and it is topologically simpler, with a greatly 
reduced number of local minima. 
The history-dependent bias prevents the system
from vistiting regions it has already explored. This term is crucial for the efficiency of the method:
for example, a Langevin dynamics in the space of the collective coordinates  would 
provide scant information on the 
underlying FES and it would be slower in locating global minima.
This has been checked by performing a Langevin dynamics for the dissociation of
a NaCl molecule in water, with the same collective variables introduced in Section 2 and adding to
eqn. (\ref{dyn}) a Gaussian noise at a temperature $T$ that is gradually reduced\cite{annealing}.  
The number of evaluations of the forces required to converge to the minimum
of the FES (i.e. the dissociated state)
is well above 1000, if the temperature is decresed with a logarithic schedule, as required in order to
ensure quasi-ergodicity in the collective coordinate space\cite{annealing1}.
A history dependent bias potential as defined in eqn. (\ref{gauss})
but applied in a regular MD simulation  without coarse graining the dynamics would be efficient in finding
escapes from the local minima, at it has been shown elsewhere\cite{Huber}, but  it would not provide 
{\it quantitative} informations about the FES. In particular, equation (\ref{FES}) holds only because
the dynamic is performed in the collective coordinates space using the derivatives of the free energy.
Hence, both the coarse-grained dynamics on the FES and the history-dependent bias are essential 
ingredients of the method, and only their combination allows an efficient and accurate determination of the FES.

Another advantage of our search algorithm is that it can be easily parallelized
over the $P$ replicas, thus  optimally exploiting present-day computer architectures and, moreover,
does not require a very accurate determination of the forces. In fact, in model calculations on an analytic
PES, we have found that the
algorithm is effective even for  noise levels as large as 20-30 \% of the forces, since the
height and the width of the gaussians is such that 
the coarse-grained dynamics explores a region of space of the size of a Gaussian several times. 
and the system is only {\it discouraged} (and not {\it prevented})
from revisiting the same spot of configuration space.
Hence, the errors in the forces tend to even out during the evolution. The dynamics (\ref{dyn}) is "self-healing",
i.e. it is in principle capable to compensate the effect of complitely wrongly located gaussians or of a wrong 
additional bias potential.
We can therefore use 
quite short runs and a small number of replicas to evaluate the forces. 
This is at variance with thermodynamic integration\cite{allen} and the blue moon 
ensemble method\cite{Constraints1,Constraints2}, where a 
very precise determination of the forces is needed and where treating multidimensional 
reaction coordinates is a daunting task. 
Other techniques, like the umbrella sampling or bias potential methods
\cite{allen,Voter1,Voter2,VandeVondele1}, allow an accurate estimate of the free energy surface in many dimensions. 
The drawback of these methods is that they
are efficient only if a good approximation for the FES is known {\it a priori}, while this is 
not required in our method.

A last advantage of the method is its ability to provide 
qualitative information on the free energy of a system in a very short time.
For example, the overall topology of a free energy surface can be determined
by very few coarse-grained dynamics steps using "large" Gaussians. Subsequently, our "qualitative"
knowledge of the FES can be improved using smaller Gaussians, eventually reducing
the dimensionality of the problem by exploiting the topological information obtained with the
large Gaussians.

MP would like to thank Yannis Kevrikidis for a very illuminating discussion on the 
coarse-grained method, and David 
Chandler for having patiently shared his 
understanding of the statistical mechanics of transition paths.
AL thanks Joost VandeVondele for many precious suggestions.

\end{document}